\documentclass[aps,prl,twocolumn,groupedaddress]{revtex4}

\usepackage{graphicx}
\begin{document}
\setcounter{page}{1}

\title{Gravitational field measurement with an equilibrium ensemble of cold atoms}

\author{Igor Kulikov}
\email[]{kulikov@jpl.nasa.gov}

\affiliation{Jet Propulsion Laboratory \\
        California Institute of Technology\\
        MS 126-347, 4800 Oak Grove Drive\\
        Pasadena, CA  91109-8099 USA}

\begin{abstract}
A new approach to the measurement of gravitational fields with an
equilibrium ensemble of ultra-cold alkali atoms confined in a cell of volume $V$ is investigated. 
The proposed model of the gravitational sensor is based on a variation
of the density profile of the ensemble due to changing of the gravitational field.  
For measurement the atomic density variations 
of the ensemble the electromagnetically induced transparency method is used.\\
PACS number(s): 39.20.+q, 03.75.Dg, 04.80.-y, 32.80.Pj, 42.50.Gy
\end{abstract}

\maketitle

Gravitational sensors operating on technologies of atom optics usually use a variety
of atom interferometry techniques for the detection of gradient of the gravitational
potential \cite{cla}-\cite{pet}. Atom interferometry requires 
elements equivalent to beam splitters
and mirrors of light interferometry \cite{ai}-\cite{mey}. 
These elements complicate the gravitational
sensor's design and impose limitations on its sensitivity. 
The objective of this work is to provide 
a scheme of gravitational sensor excluding atom interferometry technique. 
We will investigate the posibility of gravitational field measurement 
with an equilibrium ensemble of ultra-cold alkali
atoms (medium) confined in a cell of a volume $V$. The density profile 
of this medium is changed due to the
interaction of its atoms with the gravitational field. 
The gravitational field can
be measured by measuring the phase shift of an optical 
beam passing through the medium.
For measuring the density variation of the medium, 
the electromagnetically induced
transparency technique is used \cite{bol}-\cite{har2}. 
The sensitivity of the
idialized  model of the proposed gravitational sensor is comparable with
the sensitivity of modern atom
interferometric gravimeter.

Let us consider an atomic cloud confined and cooled inside a
thin-walled cell of  a small volume $V$. The technique of alkali
atom cooling is well developed and presented in a number of
publications \cite{wie}-\cite{har}. After the trapping potential is
turned off the cloud will expand filling all volume $V$. We will
consider that on this stage the atomic ensemble  is only under the
influence of a gravitational field and will neglect the heat
exchange between the ensemble of alkali atoms and the cell during
the measurement time. The field theoretical Hamiltonian of the
ensemble in the presence of the external gravitational  field 
is written as 
\begin{eqnarray}
\hat H=\int d^3 \vec x \psi^+(\vec x,t)\left[
-\frac{\hbar^2 }{2m}\nabla^2+V(\vec x)
\right]\psi(\vec x,t)
\nonumber\\
+\frac{1}{2}\int d^3 \vec x^\prime d^3 \vec x
\psi^+(\vec x,t)\psi^+(\vec x^\prime,t)U(\vec x,\vec x^\prime)
\psi(\vec x,t)\psi(\vec x^\prime,t), \label{e1}
\end{eqnarray}
where $\psi(\vec x,t)$ and $\psi^+(\vec x,t)$ are field operators,
 $V(\vec x)$ is the atom-gravitational field interaction potential, 
 $m$ is the atomic mass, and $U(\vec x,\vec x^\prime)$ is the two
body  interatomic potential \cite {dalf}. Operator $\psi(\vec x,t)$
creates an atom at the point $\vec x$ and time $t$ while
$\psi^+(\vec x,t)$ annihilates an atom. The field operators obey the
following commutation relations: $[\psi(\vec x,t),\psi(\vec
x^\prime,t)]=0$ and $[\psi(\vec x,t),\psi^+(\vec
x^\prime,t)]=\delta^3(\vec x -\vec x^\prime)$, which define the
statistics of the ensemble. We can also introduce a number operator
\begin{equation}
\hat N=\int d^3 \vec x\psi^+(\vec x,t)\psi(\vec x,t) \label{e2}
\end{equation}
with the eigenvalues interpreted as the number of the atoms.
For further analysis we will
use the grand partition function of the atomic medium
\begin{equation}
Z={\rm Sp}
\left[\exp{(\beta(\mu \hat N-\hat H))}
\right], \label{e3}
\end{equation}
where $\mu$ is the chemical potential, and $\beta=1/k_BT$ is the inverse temperature
($k_B$ is the Boltzmann constant).
 If  atoms of the medium are under the influence of only a gravitational field, then the potential of
 interaction is given by the classical equation $V(\vec x)=m\Phi(\vec x)$, where $\Phi(\vec x)$
 is the Newtonian gravitational potential.
 Inserting (\ref{e1}) and (\ref{e2}) into (\ref{e3})
 and neglecting interatomic interactions \cite{grav} we obtain the
 grand partition function in the form
\begin{equation}
Z=\prod_i \left[1-\exp{(-\beta(\epsilon_i-\mu))}
\right]^{-1}, \label{e4}
\end{equation}
where $\epsilon_i$ is the energy of a boson in $i$-th state. Statistical and thermodynamical properties
of the medium can be investigated with the help of a grand thermodynamic potential given by
$\Omega=-(1/\beta)\ln Z$.
Using the equation (\ref{e4}) we obtain the expression for the thermodynamic potential in the form
\begin{equation}
\Omega=(1/\beta)\sum_i \ln
\left[1-\exp{(-\beta(\epsilon_i-\mu))}
\right]. \label{e5}
\end{equation}
In quasi-classical regime we can replace in the equation (\ref{e5}) the summation over single
atomic states by the integration
$ \sum\nolimits_i \rightarrow s\int(2\pi \hbar)^{-3}d \Gamma$
, where $d \Gamma=d^3 \vec p d^3 \vec x$, and $s$ is the number of the possible spin
states of the atom. The atomic energy in this case can be represented by a sum of kinetic
and potential energies $\epsilon_i={\vec p~}^2/2m +m\Phi(\vec x)$
 and the expression for the grand thermodynamic potential will be
\begin{equation}
\Omega=(1/\beta)\int \frac{s d\Gamma}{(2\pi \hbar)^3}
 \ln
\left[1-z(\Phi)\exp{(-\beta p^2/2m)}
\right], \label{e6}
\end{equation}
where the gravity dependent fugacity $z$ is given by the equation
\begin{equation}
z(\Phi)=\exp\left[
\beta (\mu-m\Phi)\right]
. \label{e7}
\end{equation}
The grand thermodynamic potential (\ref{e6}) can be written in the form
$ \Omega=\int d^3 \vec x \omega(\vec x)$
where the density of
grand thermodynamic potential  is given by the equation
\begin{equation}
\omega=(1/\beta)\int \frac{s d^3\vec p}{(2\pi \hbar)^3}
 \ln
\left[1-z(\Phi)\exp{(-\beta p^2/2m)}
\right]. \label{e8}
\end{equation}
The average density of atoms in the medium is found as $N=-(\partial\omega /\partial \mu )$
and has the following form
\begin{equation}
N=\int \frac{ d^3\vec p}{(2\pi \hbar)^3}
\frac{s}{z^{-1}(\Phi)\exp(\beta p^2/2m))-1}.
\label{e9}
\end{equation}
The integration of (\ref{e9}) yields
\begin{equation}
N\lambda^3_{dB}=g_{3/2}(z),
\label{e10}
\end{equation}
where $\lambda_{dB}=\sqrt{2\pi \hbar^2\beta/m}$ is the thermal de Broglie wavelength
and $g_{3/2}(z)$ is a function defined by the equation
\begin{equation}
g_k(z)=\sum^\infty_{n=1} \frac{z^n}{n^k}. \label{e11}
\end{equation}
The functions $g_k(z)$ obey the recursion relations 
\begin{equation}
z\frac{d}{dz}g_k(z)=g_{k-1}(z) \label{e12}
\end{equation}
and have the following integral representations
\begin{equation}
g_k(z)=\frac{2^{2k-1}}{\sqrt \pi}\frac{\Gamma(k-1/2)}{\Gamma(2k-1)}
\int^\infty_0 dx\frac{x^{2k-1}}{z^{-1}\exp(x^2)-1}, \label{e13}
\end{equation}
where $\Gamma(k)$ is the gamma function. The functions $g_k(z)$ have
singularity at $z=1$ , which leads to BEC, moreover, the series
(\ref{e11}) converges only for $k >1$. The equation (\ref{e10})
describes the relation between the density of particles $N$,
temperature $T$ and the gravity dependent fugacity $z$. The
variation of the atomic density of the medium is found from
(\ref{e10}) in the form 
\begin{equation}
\delta N=-\beta m N_0 F(z_0)\delta \Phi,
\label{e14}
\end{equation}
where $F(z)$ is given by the equation $F(z)=g_{1/2}(z)/g_{3/2}(z)$.
In the equation (\ref{e14}) the atomic density variation $\delta N=N-N_0$,
and the variation of gravitational potential $\delta\Phi=\Phi(\vec x)-\Phi({\vec x}_0)$
are defined between points $\vec x$ and ${\vec x}_0$.
Fugacity $z_0$ at point ${\vec x}_0$ is related to $N_0$
and temperature $T$ of the medium by the equation (\ref{e10}).
Let us select the coordinate system $\{x^1,x^2,x^3\}$,
where the vector of gravitational acceleration is $\vec g=\{0,0,-g \}$.
In this coordinate system the variation of gravitational potential
can be written as $\delta \Phi=gY$,
where $Y=(x^3-x^3_0)$ is the displacement between two points along $x^3$-axis.

The equation (\ref{e14}) can be used for the analysis of the
influence of gravitational field on the medium consisting of
non-interacting alkali atoms in the cell of volume $V$. Let us
rewrite it in the form 
\begin{equation}
\delta N=-N_0 \beta m YF(z_0)g.
\label{e15}
\end{equation}
As follows from this equation, the gradient of gravitational
potential causes the variation of the atomic density of the medium
in the cell that can be detected with electromagnetically induced
transparency (EIT) technique in the combination with the method of
optical interferometry (Fig. 1).

\begin{figure}
\includegraphics[width=0.6\columnwidth,angle=-90]{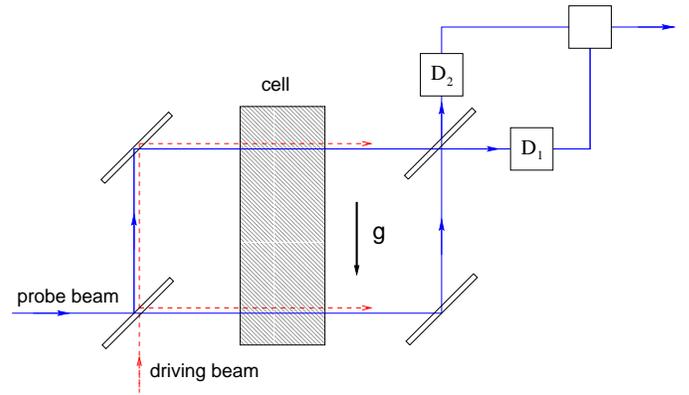}
\caption{Mach-Zehnder interferometer detects the phase shift of the
probe beams after they pass through the cooled atomic medium placed
in the cell of the volume $V = 1~cm^3$.   The gradient of
gravitational potential causing the phase shift is given by an
arrow. The cell has the length $L$, and the upper and the lower
beams of the interferometer are separated by the distance $Y$.
Dashed arrows correspond to driving fields.\label{fig:LAT96}}
\end{figure}

Let us consider EIT effect in application to the medium in
gravitational field by taking into account the internal structure of
bosons. We will assume that each boson of the medium is a
three-level system with one upper (a) and two lower levels (b) and
(c). To obtain EIT conditions, two optical (driving and probe)
fields are used. Let the driving field of Rabi frequency $\Omega$
couple the levels (a-c) whereas the probe field of frequency
$\omega$ couple the levels (a-b). The driving field provides an
interference of different possible absorption pathways producing no
absorption effect \cite {har3}. As the result, the medium becomes
transparent to a probe field and the real part of susceptibility
$\chi^ \prime$ of the probe frequency near the resonant point will
be 
\begin{equation}
\chi^\prime=-\frac{\wp^2}{\hbar \varepsilon_0}\frac{\Delta}{\Omega^2}N,
\label{e16}
\end{equation}
where $\wp$ is the $a \rightarrow b$ transition matrix element, $N$ is the atomic density,
and $\Delta_{ab}=\omega_{ab}-\omega$ is
the atom-probe field detuning \cite {sc}. The dispersive properties of the medium
are defined by refractive index $n=1+\chi^\prime /2$, which is written in the form
\begin{equation}
n=1-\frac{\wp^2}{2\hbar \varepsilon_0}\frac{\Delta}{\Omega^2}N.
\label{e17}
\end{equation}
As follows from Fig.1, the interferometric phase shift $\phi$ of probe beams is
a function of the variation of refractive index $\delta n$ of the medium
\begin{equation}
\phi=\frac{2 \pi L}{\lambda}\delta n,
\label{e18}
\end{equation}
where $\lambda$ is the wavelength of the probe field, and $L$ is the
length of the cell. Interaction of atoms of the medium with
gravitational field changes the density profile of the medium. The
density variation in the medium leads to variation of the refractive
index along the gradient of gravitational field. When two separated
probe beams pass the medium at points of different gravitational
potential, they gain the phase shift depending on the gradient of
the gravitational potential. Therefore the gradient of the
gravitational potential can be measured by measuring the phase shift
of probe beams in the interferometer.

From the equations (\ref{e15}), (\ref{e17}) and (\ref{e18}) we obtain the relation between the
phase shift and the gravitational acceleration. The variation of the refractive
index is found from the equation (\ref{e17}) as
\begin{equation}
\delta n=\left [\frac{3}{8 \pi^2} \frac{\lambda^3 \gamma \Delta}{\Omega ^2}
\right] \delta N,
\label{e19}
\end{equation}
where $\gamma=\wp^2 \omega ^3/(6\pi \hbar \varepsilon_0 c^3)$
is the radiative decay from the upper level. Then the phase shift of
the two beams in the interferometer will be
\begin{equation}
\phi=\left [\frac{3}{4 \pi} \frac{L \lambda^2 \gamma \Delta}{\Omega ^2}
\right] \delta N.
\label{e20}
\end{equation}
Inserting the equation (\ref{e15}) into the equation (\ref{e20}) we get
\begin{equation}
\phi=\left [\frac{3}{4 \pi} \frac{A \lambda^2 \gamma^2 x N_0 m F(z_0) }{ k_B T \Omega ^2}
\right] g,
\label{e21}
\end{equation}
where $A=LY$ is the area of the side surface of the medium, 
and $x=\Delta/\gamma$ is the dimensionless
 detuning scale. The limiting sensitivity of the device is defined by the equation
\begin{equation}
\delta g=\left [\frac{3}{4 \pi} \frac{A \lambda^2 \gamma^2 x N_0 m F(z_0) }{ k_B T \Omega ^2}
\right]^{-1} \delta \phi.
\label{e22}
\end{equation}

For numerical demonstration of the limiting sensitivity of this gravitational sensor,
 we will assume that the driving field is strong and $\Omega \approx \gamma$ \cite {fsc}.
 Taking into account that $F(z) \ge 1$,
 and putting for simplicity $xF(z) \approx 1$ we can rewrite
 the equation (\ref{e22}) in a simple form \cite {ik}
\begin{equation}
\delta g=\left [\frac{4 \pi}{3} \frac{k_B}{ A \lambda^2 N_0m}
\right] T \delta \phi. \label{e23}
\end{equation}
For a wavelength $\lambda=500~{\rm nm}$, atomic density
$N_0=10^{12}~{\rm cm^{-3}}$, atomic mass $m=10^{-26}~{\rm kg} $, and
the side area of  the cell  $A=1~{\rm cm^2}$, we find that the
coefficient of the equation (\ref{e23}) in square bracket 
is of the order of ten and, therefore, one can simplify the equation (23) as
$\delta g \approx 10 T \delta \phi$. Assuming the temperature of the gas is
about $T= 1~{m\rm K}$ and computing the sensitivity of the interferometer
with the equation  $\delta \phi = \sqrt{\hbar \omega/P_p \tau}$ for
$P_p=1~m$W probe beam power and $\tau=1$ second of measurement time, we find
that the minimal detectable gravitational field changing is
 $\delta g \approx  10^{-11}~{\rm g}$.
As follows from the equation (\ref{e23}), the important factor contributing
to the sensitivity of the sensor is the side area of the cell $A$.
The volume $V$ of the cell contributes to the sensitivity due to
changing the density of atoms $N_0$. Therefore probe beam separation
$Y$ (height of the cell), length of the cell $L$ and its thickness
can be adjusted based on optical interferometer design.

In conclusion, the numerical analysis has demonstrated  
that cold-atom EIT gravitational sensor
operating with an equilibrium ensemble of cold alkali atoms 
confined in one cubic centimeter volume allows us to measure
gravitational acceleration with high sensitivity. The sensitivity of 
the proposed gravitational sensor operating 
with  $m$K temperature alkali medium is about $10^{-11}{\rm g}$ which is
higher for the same operational time then that of modern
cold-atom gravimeters operating on atom interferometry effect.
Lowering the temperature of alkali medium to $\mu$K regime, which is
the operational temperature of the atom-cloud gravimeter, one can reduce the number
of atoms from $N_0=10^{12}$ to $N_0=10^{9}$ ${\rm cm^{-3}}$ keeping the sensor's sensitivity 
$10^{-11}{\rm g}$. 
As follows from the above analysis, obtaining high sensitivity requires high atomic density 
and precise control of cloud temperature. The better we control temperature the
longer is the time of phase measurement.
The idialized model which has been investigated here does not include thermal exchange between 
alkali cloud and the cell. We assumed the temperature is constant during the measurement time $\tau$.
 Thermal interaction shortens the time of measurement and lowers
the sensitivity of the sensor. 
A possible way to reduce the effect of thermal exchange on the sensor's sensitivity 
is to compensate for the raise of the temperature 
by experimentally defined (or modelled) function $T(t)$, $t \in \tau$ 
in the equation (\ref{e23}). 
Due to time dependence of the temperature of the ultra-cold cloud the proposed model operates in discrete regime. 
For each consequent measurement the cell must be refilled with a new collection of atoms.

\bigskip
\begin{acknowledgments}
This work was done at the Jet Propulsion Laboratory, California
Institute of Technology under a contract with the National
Aeronautic and Space Administration. 
\end{acknowledgments}


\begin{thebibliography}{99}
\bibitem{grav} Atomic collisions don't change gravity-dependent atom-density profile and
can be omitted at this step of computation.
\bibitem{ik}    The fugacity $z$, given $N$ and $T$, is found from the equation (\ref{e10}).
For small $z$ the function $F(z)$ has the limit $F(z \ll 1)=1$. 
This limit corresponds to Boltzmann statistics, and the particle density profile is defined by
 the barometric equation. At low temperatures, due to divergence of $g_{1/2}(z)$,
 the function $F(z)$ will grow at $z \rightarrow 1$ and the effect of Bose-Einstein statistics
 will take place. For the numerical demonstration it was assumed that $xF(z)=1$,
 though at low temperatures this product will be large.
\end{thebibliography}
\end{document}